\documentclass{PoS}
\usepackage{lineno}
\title{Performance of the SoLid Reactor Neutrino Detector}

\ShortTitle{Performance of the SoLid Reactor Neutrino Detector}

\author{\speaker{Luis Manzanillas}\thanks{On behalf of the SoLid  collaboration}\\
        LAL, Univ Paris-Sud, CNRS/IN2P3, Universit\'e Paris-Saclay, Orsay, France\\
        E-mail: \email{manzanillas@lal.in2p3.fr}}

\abstract{The SoLid  collaboration is currently operating a 1.6 tons neutrino detector near the Belgian BR2 reactor, 
with main goal the observation of the oscillation of electron antineutrinos to previously undetected 
flavor states. The highly segmented SoLid  detector employs a compound scintillation technology based on 
PVT scintillator in combination with a $^{6}$LiF:ZnS(Ag) screens containing $^{6}$Li isotopes. The experiment
has demonstrated a channel-to-channel response that can be controlled to the one percent level, and energy resolution of better than 14\% at 1 MeV, and a 
determination of the interaction vertex with a precision of 5 cm.
In this contribution we highlight the the current performance and stability of the full-scale system. The  in-situ calibration results of the detector with various radioactive sources are discussed as well.}

\FullConference{
European Physical Society Conference on High Energy Physics - EPS-HEP2019 -\\
			10-17 July, 2019\\
			Ghent, Belgium}

\begin{document}

\section{Introduction}
Several anomalies in short baseline $\nu$ experiments have been observed in the previous years. In particular the Gallium Anomaly and the Reactor Antineutrino Anomaly (RAA) both show a deficit of neutrinos with respect to the expectations at the 3$\sigma$ level. The most recent evaluations of these anomalies suggest that the existence of a light sterile neutrino state with a mass at the eV scale could account for such deficits \cite{Gariazzo:2017fdh,Dentler:2018sju}. In addition, recent reactor $\nu$ experiments have also found a distortion in the $\overline{\nu}_e$ energy spectrum, which is known as the ``5 MeV bump''. The origin of these anomalies remains unknown and the consensus among the neutrino community is that new data from short baseline reactor experiments is needed in order to clarify these issues. These experiments can only be realized at the surface level in the vicinity of a nuclear reactor plant, which implies high backgrounds induced by cosmic rays and by the reactor itself. 

The expected signature of a sterile $\nu$ is the deformation of the $\overline{\nu}_e$ energy spectrum as function of both the  $\overline{\nu}_e$'s energy and the distance traveled by them from the source to the detection point. Thus a good energy and space resolutions are of paramount importance for detecting this imprint. In addition, such experiments should provide as much tools as possible to discriminate the high backgrounds present at the surface level.  
In this context the SoLid collaboration is operating a 1.6 tons $\overline{\nu}_e$ detector near the Belgian BR2 research reactor. The goals of SoLid are to look for an energy and space oscillation pattern at shord baselines in order to confirm or reject the sterile $\nu$ hypothesis as the origin of the RAA, and provide a new measurement of the $^{235}$U reactor fuel.

\section{SoLid  technology}
\begin{figure}[h]
\centering
\begin{tabular}{c c}
  \includegraphics[width=0.3\linewidth]{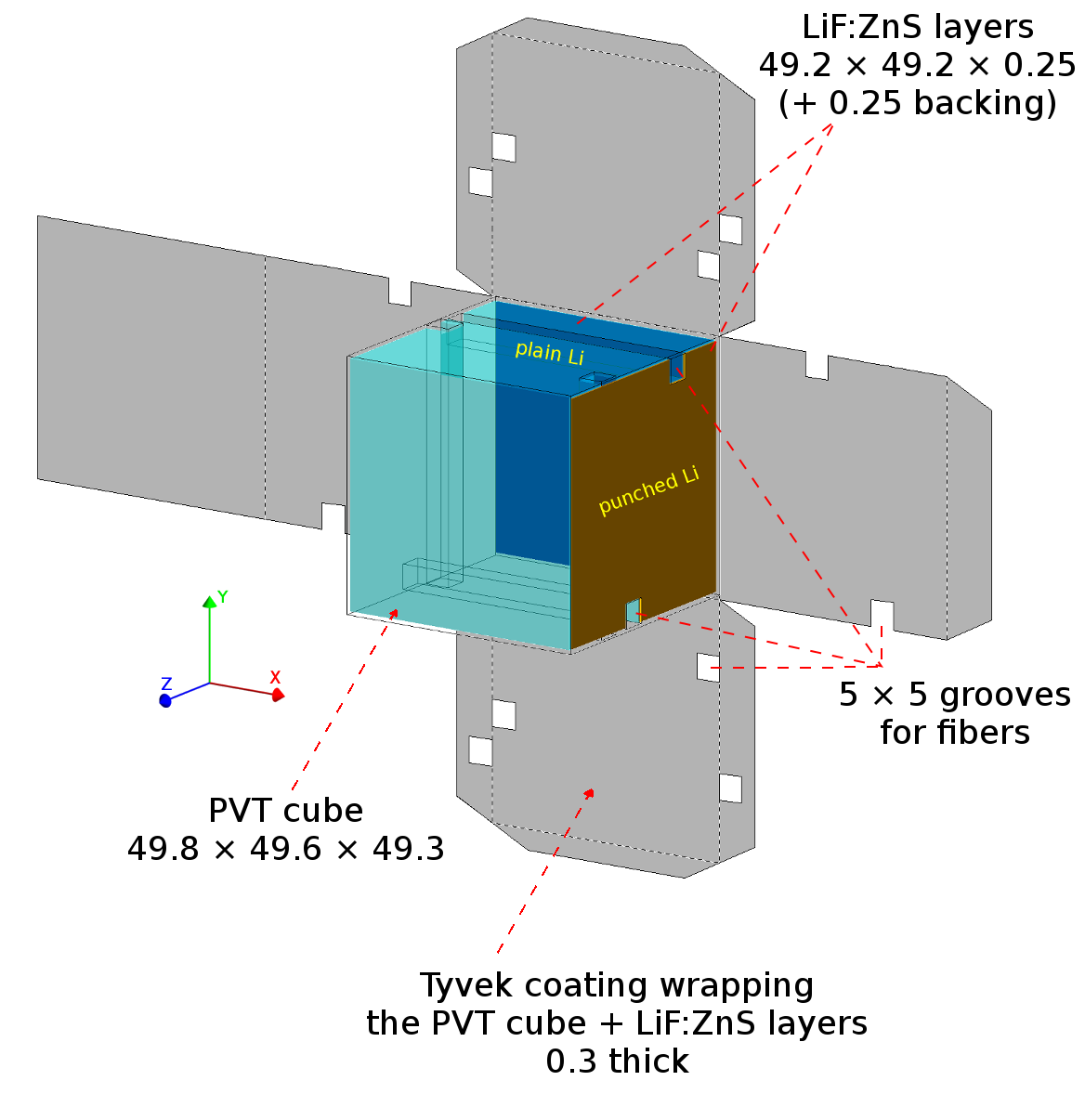} & \includegraphics[width=0.7\linewidth]{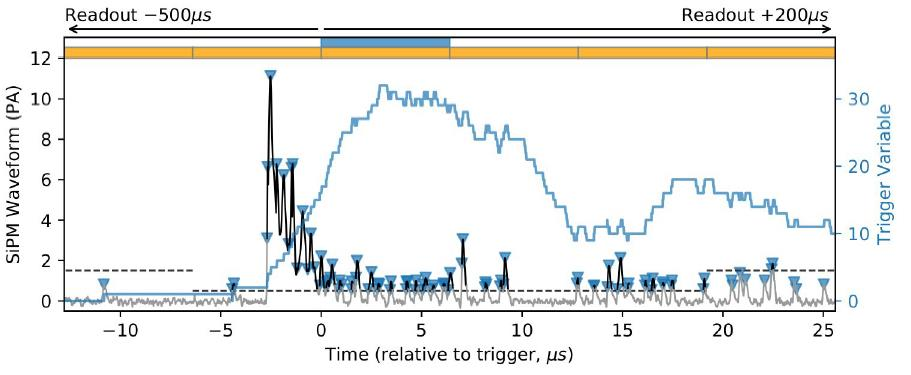}
\end{tabular}
\caption{Left : A SoLid cube. Right: Neutron trigger in SoLid.} 
\label{fig:nTrigger}
\end{figure}

Reactor $\overline{\nu}_e$'s are detected via the Inverse Beta Decay (IBD) process : $\overline{\nu}_e + p \rightarrow e^{+}+n$,
whose signature, are prompt-delayed signals correlated in time and space. To detect this signature the SoLid collaboration has developed a new detector concept for $\overline{\nu}_e$ detection.  SoLid employs a hybrid detector composed of PVT cubes of 5$\times$5$\times$5 cm$^{3}$ dimensions, combined with two screens of $^{6}$LiF:ZnS(Ag) in two faces of each cube as shown in figure \ref{fig:nTrigger} (Left). This ensemble is then wrapped with Tyvek in order to make optical isolation and constitutes a detector cell. The full detector is composed of 12800 cells that are distributed in 5 modules of 10 planes each, and each plane is made of an array of 16$\times$16 cells. Finally light signals are readout using a network of wavelength shifting fibers (64 per plane) coupled to one end to a SiPM and to the other end to a mirror.

PVT is used as target for the IBD process and as calorimeter for the $e^{+}$ interaction and detection, which provides the required information to assess the $\overline{\nu}_e$ initial energy. On the other hand, neutrons are captured and detected on the  $^{6}$LiF:ZnS(Ag) screens after thermalization in the PVT. Most of the neutrons are captured on the $^{6}$Li nuclei, wich will break up depositing about 4.7 MeV in the ZnS scintillator. Time constants for light emission of PVT (ES) and ZnS (NS) are of the order of ns and $\mu$s respectively. Hence neutron captures on the $^{6}$Li will produce much longer waveforms composed of several pulses of light as shown in figure \ref{fig:nTrigger}(Right). SoLid exploits the characteristics of these NS signals for trigger and PSD discrimination \cite{Abreu:2018njy}.

\section{Construction, Quality Assurance and  Commissioning}
For the construction of the full detector about 13000 PVT cubes were received and underwent a cleaning process, being manually washed, weighted, wrapped and stacked in hermetic plastic boxes before being mounted in planes. Detailed information about weight, number of batch, Li content, and all relevant information was stored in a dedicated database. In order to guarantee and validate the minimal requirements of light yield and neutron detection efficiency for the SoLid physics program, all the planes were qualified before being assembled into modules. To this end, an automated robotic  system called CALIPSO was developed, which allowed for a preliminary calibration of each plane \cite{Abreu:2018ekw}.  During this QA process a problem with a batch of Li screens was identified. This batch of screens were loaded with only 50\% of the expected $^{6}$Li content and therefore the neutron detection efficiency was too low in the cubes mounted with screens from this batch. All cubes assembled with this problematic Li screens had to be replaced. In addition, some other minor problems as bad couplings and unresponsive SiPMs were also identified and fixed. Finally, a light yield and a neutron reconstruction efficiency larger than 60 PA/MeV and 60\% respectively were estimated, which exceeded the initial SoLid requirements. Thus the first modules were deployed at BR2 at the end of 2017 and the full detector was commissioned on February 2018. After some trigger and data rate optimizations the detector is taking data in physics mode in stable conditions since May 2018.  

\section{Detector Operation and Performances}
The SoLid detector is operated running three types of trigger in parallel. A first random trigger is used to readout non zero-suppressed waveforms of the whole detector at 1 Hz. The main purpose of this trigger is for monitoring the stability of the SiPMs. A threshold trigger asking for 2 MeV in XY coincidence of two fibers is also used. This trigger is mainly used as muon and high electromagnetic event tagger. Finally a trigger for $\overline{\nu}_e$ detection is employed using the NS signals. This NS trigger is based on an algorithm that counts the number of peaks above a certain threshold in a rolling time window. Once that a NS is triggered, the readout of all what happened in the previous 500 $\mu$s and the next 200 $\mu$s in  the triggered plane plus +/- 3 planes is performed. In this way an unbiased prompt detection can be realized, and a high IBD efficiency can be achieved.  

\begin{figure}[h]
\centering
\includegraphics[width=0.7\textwidth,trim={0cm 0.2cm 0cm 0.8cm},clip]{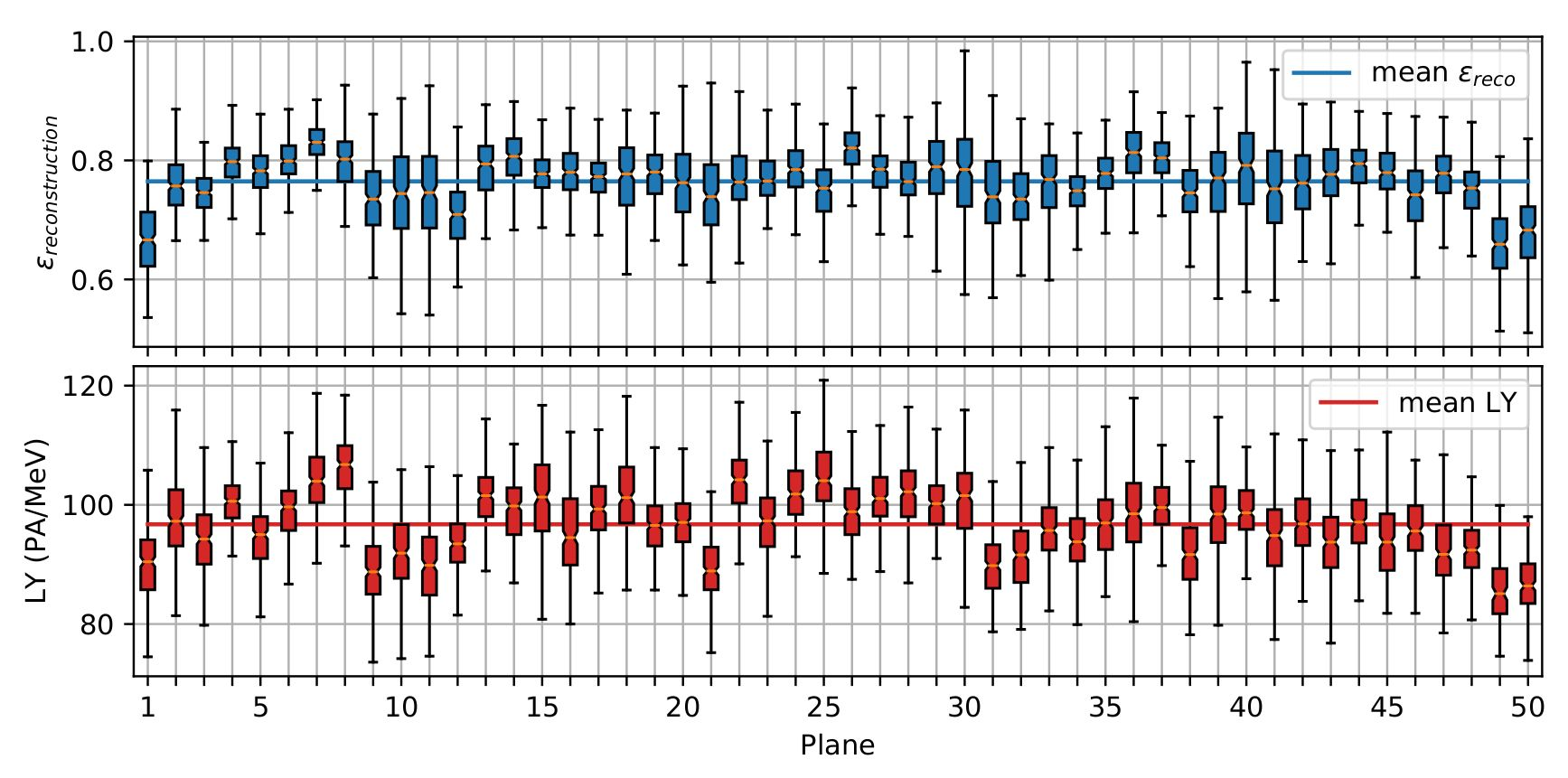}
\caption{Light yield and neutron reconstruction efficiency of the 50 planes of the SoLid detector.}
\label{CalibResults}
\end{figure}

In order to determine the light yield and neutron detection efficiency with the final detector settings, an automated calibration system called CROSS was designed and constructed. This system allows to insert radioactive sources between two modules for calibration purposes. The sources are placed in 9 different positions per gap, which has been optimized to have a distance source to farest detector cell smaller than 35 cm.  $^{22}$Na (511 keV and 1270 keV $\gamma$'s) and AmBe (<4.2 MeV> $n$'s) sources are used every reactor OFF period for light yield and $n$ detection efficiency calibration. The calibration of the full detector last one day for  $^{22}$Na and three days for AmBe. In addition, dedicated calibration campaigns have been performed with other gamma sources  for energy linearity studies. These results demonstrated an energy linear response in the region of interest for reactor $\overline{\nu}_e$'s. On the other hand $^{252}$Cf (<2 MeV> $n$'s) has been also used, which provides a better control of systematic uncertainties on the neutron detection efficiency. Thus, in average a light yield of about 96 PA/MeV and a $n$ reconstruction efficiency larger that 70\% have been measured with a dispersion smaller than 10\%, which evidence an excellent performance of the detector (see figure \ref{CalibResults}). 
\section{Conclusions}
The SoLid collaboration has developed a new detector concept for reactor $\overline{\nu}_e$ detection. The construction of a 1.6 tons detector has been successfully completed on February 2018. The detector is taking data in physics mode in stable conditions since May 2018. The calibration results indicate an excellent detector performance, with a light yield and a neutron reconstruction efficiency larger than 70 PA/MeV and 70\% respectively. In addition, a linear energy response in the region of interest for reactor $\overline{\nu}_e$ has been demonstrated. The analysis of a first unblinded data sample is going and the first physics results are expected soon.

\end{document}